\documentclass[journal]{IEEEtran}
\usepackage{graphicx,epsfig,amsmath,amssymb,bm}
\usepackage{hyperref}
\hypersetup{urlcolor=cyan}

\usepackage{amssymb,balance}
\usepackage{amsmath}
\usepackage{amsthm} 
\usepackage{amssymb}
\usepackage{algorithmicx}
\usepackage{algpseudocode}
\usepackage{cite}
\usepackage{url}
\usepackage{bbm}
\usepackage{subcaption}
\usepackage{float}
\usepackage{tabularx}
\usepackage{mathtools}
\usepackage{empheq}
\usepackage{multirow}
\usepackage{verbatim}
\usepackage[ruled,linesnumbered,noend]{algorithm2e}
\usepackage{tikz}
\usepackage{microtype}
\usepackage[capitalise,noabbrev]{cleveref}
\include{defn}

\newcommand{\beqa}{\begin{eqnarray}}
	\newcommand{\eeqa}{\end{eqnarray}}

\newcommand{\ale}[1]{{\color{blue} [Ale: #1]}}
\ifCLASSINFOpdf

\else

\fi

\begin{document}
\title{Uncovering Load-Altering Attacks Against $N-1$ Secure Power Grids: A Rare-Event Sampling Approach}
\author{Maldon Patrice Goodridge, Subhash~Lakshminarayana~\IEEEmembership{Senior Member, IEEE}, and Alessandro Zocca \vspace{-0.35in}
\thanks{M.P. Goodridge is with Global Development Initiatives, UK, (email: \texttt{gdi.uklimited@gmail.com}). S.~Lakshminarayana (Corresponding author) is with the School of Engineering, University of Warwick, Coventry, UK, CV47AL (email: \texttt{subhash.lakshminarayana@warwick.ac.uk}). A.~Zocca is with the Department of Mathematics, Vrije Universiteit Amsterdam, NL (email: \texttt{a.zocca@vu.nl}). A preliminary version of the paper was published at PMAPS 2022 \cite{LaksRareEvent2022}.
}}
	
		

	

	\maketitle

\begin{abstract}
Load-altering attacks targetting a large number of IoT-based high-wattage devices (e.g., smart electric vehicle charging stations) can lead to serious disruptions of power grid operations. In this work, we aim to uncover spatiotemporal characteristics of LAAs that can lead to serious impact. 
The problem is challenging since existing protection measures such as $N-1$ security ensures that the power grid is naturally resilient to load changes. Thus, strategically injected load perturbations that lead to network failure can be regarded as \emph{rare events}. To this end, we adopt a rare-event sampling approach to uncover LAAs distributed temporally and spatially across the power network.  The key advantage of this sampling method is the ability of sampling efficiently from multi-modal conditional distributions with disconnected support. Furthermore, we systematically compare the impacts of static (one-time manipulation of demand) and dynamic (attack over multiple time periods) LAAs. 
We perform extensive simulations using benchmark IEEE test simulations. The results show (i) the superiority and the need for rare-event sampling in the context of uncovering LAAs as compared to other sampling methodologies, (ii) statistical analysis of attack characteristics and impacts of static and dynamic LAAs, and (iii) cascade sizes (due to LAA) for different network sizes and load conditions. 
\end{abstract}

	\IEEEpeerreviewmaketitle
	
\section{Introduction}

The growing penetration of internet-of-things (IoT) enabled high-wattage appliances (e.g., electric vehicle charging stations) may create new malicious attack opportunities against power grids. \emph{Load-altering attacks} (LAAs) are large-scale load fluctuations that can be generated by changing the operational settings of IoT-based load devices under a botnet-type attack. Such LAAs can pose a serious threat to power grid operations \cite{Mohsenian_LAA2011, Dabrowski2017, Soltan2018, HuangUSENIX2019, AminiLAA2018, LakshIoT2021}. 
LAAs can be classified into two categories: (i) \emph{static} LAAs (S-LAAs), which refers to a sudden one-time manipulation of power grid demand, and (ii) \emph{dynamic} LAAs (D-LAAs), which refers to a series of load alterations over time. 

Analyzing the impact of LAAs on power grid operations is an important step toward developing measures to enhance system resilience. It was shown \cite{Dabrowski2017, Soltan2018} that the imbalance caused due to the mismatch between the supply and demand under S-LAAs can lead to unsafe frequency excursions causing line outages, generator trips, and/or increase  the grid's operational costs. D-LAAs, on the other hand, do not have an instantaneous impact, but however, can potentially destabilize the power grid's frequency control loop over a period of time \cite{AminiLAA2018}. In \cite{HuangUSENIX2019}, the authors analyzed the LAA impact under more realistic operational regimes including protection measures such as under-frequency load shedding, over-frequency protections, etc., and showed that large-scale LAAs can still cause islanding and generator disconnections.
Subsequent work has investigated the impact of spatial (i.e., location of the attack within a power grid network structure) and temporal (i.e., the time of the attack at different load conditions) aspects of the attack \cite{LakshIoT2021, Ospina2021, LakshCOVID2022}. 
A framework to determine the locations of the grid (i.e., victim buses) from which an attacker can launch the most impactful attacks was formulated in \cite{LakshIoT2021} using the theory of second-order dynamical systems. Furthermore, reference  \cite{LakshCOVID2022} showed that low inertia conditions due to high penetration of renewable energy resources can exacerbate the consequences of LAAs.



Recent work has also focused on defending against LAAs. These measures can be classified into (i) offline strategies, such as the optimal deployment of protection-enabled loads \cite{AminiLAA2018} or finding generator operating points that can prevent the destabilizing effects of LAAs \cite{SoltanTNSE2020}, and (ii) online measures such as real-time detection and localization of LAAs by monitoring the load data \cite{YoussefDetect2022} or frequency/phase angle fluctuations from phasor measurement units following an LAA \cite{AminiIdentification2019, lakshminarayana2021datadriven}. Reference \cite{ChuMitigationTSG2022} proposes an online strategy to mitigate the destabilizing effects of LAAs by redispatching fast-acting inverter-based resources.
 

Despite the growing work on LAAs, existing literature lacks a framework to understand the full extent of the consequences of LAAs in realistic power grid conditions and a formal comparison of different attack strategies. 
In this work, we address this critical research gap. Firstly, we identify specific LAAs that can lead to power grid failures, even when the system is scheduled for $N-1$ contingencies. Secondly, we extensively compare the impact of two types of LAAs (S-LAAs and D-LAAs) to better understand their relative importance in power grid stability. We elaborate on the importance and challenge of these two aspects in the following.



First, $N-1$ scheduling policies provide resilience to power grids against various contingencies, including LAAs. To the best of our knowledge, the majority of the work in this area, with the exception of \cite{HuangUSENIX2019} and our precursor work \cite{LaksRareEvent2022}, has ignored $N-1$ scheduling. However, \cite{HuangUSENIX2019} considers only a few specific attack scenarios (e.g., ``IoT demand attack even distributed across all load points'') and only a few power grid operational conditions (e.g., system load). Again to the best of our knowledge, there is no work that has considered different distributions of the attack across the victim nodes and grid conditions.  
However, discovering risk-posing LAAs is challenging since $N-1$ scheduling implies that component disconnections due to LAAs are extremely rare. To overcome this challenge, we employ a specialized sampling approach known as the \emph{skipping sampler} (see \cite{Moriarty2019}), which has been used in the literature to obtain samples of high-impact events of low likelihood in low-inertial power systems \cite{goodridge2021}. The application in our case is significantly more challenging, however, since simulating D-LAAs requires sampling multiple load changes at each node in the network, which can result in a challenging high-dimension sampling problem. To address this, we extend the skipping sampler algorithm to utilize multiple skipping trajectories per proposal instead of the single trajectory used in the works cited. The points sampled on each trajectory can be understood to represent the nodal load changes applied to the network. Thus, D-LAAs comprised of $N$ load changes over time can be generated from $N$ skipping trajectories.



 
Second, as noted before, S-LAAs and D-LAAs are qualitatively different attack strategies. However, it remains uncertain which of the two strategies has a greater impact. To address this crucial question, we leverage the skipping sampler framework introduced earlier. To this end, we introduce a realistic and generalized model of D-LAAs. In particular, the original D-LAA proposed in \cite{AminiLAA2018} involves (i) continuous change in the system load with infinitesimally small intervals between successive attacks, and (ii) follows a pre-defined pattern, namely, the attack increases the load when the system frequency drops below the setpoint and decreases the load when the system frequency is above the set-point (thus worsening the generation-load imbalance at all times). Both these assumptions have limitations. Continuous load changes are unrealistic considering the attacker's limited access and load response times. Moreover, the predefined attack pattern fails to capture other sophisticated strategies that attackers might employ to destabilize the system. We generalize both these aspects by (i) considering a discretized attack model with a periodic and non-zero interval between successive attacks, and (ii) independently sampling load attacks across time and space. By varying the interval between the attack, we can transition from dynamic (short intervals between attacks) and static attacks (long interval between attacks).

We address these two research questions by performing extensive simulations using the skipping sampler. We simulate the power grid's transient dynamics using a third-order model that accounts for both the frequency as well as the voltage dynamics \cite{schm2014}. We treat the cumulative magnitude of LAA load changes that perturb the dynamical system as random variables whose realisations are sampled independently at each target node according to a log-normal distribution\footnote{Note that the proposed framework works independently of the chosen underlying cyber-attack distribution.}, which was assumed to reflect a realistic magnitude occurrence distribution based on the empirical data from data breaches \cite{Edwards2016}.
We perform simulations including appropriate emergency responses (ERs), such as generation/load shedding, etc., for different system load conditions (evening, night, etc.), and various degrees of network vulnerability to load-altering attacks.



Our results provide several novel insights into different strategies that an adversary can potentially adopt to cause a grave impact on the system by manipulating large-scale demand. The results show that adversaries employing D-LAA strategies can induce the disconnection of critical network components when manipulating a smaller magnitude of loads in the network as compared to S-LAAs. Furthermore, while S-LAA strategies cause the most damage to the power grid during low-demand periods, the system is particularly vulnerable to D-LAAs during the peak demand period. We also identify the dominant failure modes in these different load conditions (in terms of emergency response that is activated most frequently). 

This paper significantly extends our preliminary work \cite{LaksRareEvent2022}, providing an in-depth extended analysis. Specifically, the contributions of this paper can be summarized as follows: 
\begin{itemize}
    \item We propose a rare-event sampling approach to uncover LAAs that lead to power system emergency actions in $N-1$ secure power grids. We compare its performance  with several other sampling methodologies and show that the skipping sampler identifies significantly more attack strategies (in terms of load attack distributions over the victim nodes and attack time window) that can lead to network component disconnections. 
    \item Our previous analysis \cite{LaksRareEvent2022} only considered S-LAAs, but in this paper, we expand our investigation to include D-LAAs. Unlike S-LAAs, these dynamic attacks enable adversaries to exploit both spatial and temporal distributions of load changes to elicit network disruptions. However, analyzing the risks posed by D-LAAs comes with increased sampling complexity. To address this challenge, we extend the functionality of the skipping sampler algorithm to include multiple, simultaneous skipping trajectories.
    \item We exhaustively compare the impacts of S-LAAs and D-LAAs in terms of (i) the proportion of vulnerable loads on the network, (ii) the average magnitude of each load change prior to the first disconnection event, (iii) the interval between attacks, and (iv) the diurnal load balance.
\end{itemize}

 
	
The rest of the paper is organised as follows. \Cref{sec:Prelim} introduces the system model; \Cref{sc:statmodel} presents the statistical model and details of the rare-event sampling method. \Cref{sec:Sims} discusses the simulation results and \Cref{sec:Conc} concludes. 
	
\section{System model}\label{sec:Prelim} 
The power system model for rare-event sampling is based on the \textit{3rd-order model} for generators \cite{kundur1994power}. In addition, our mathematical model accounts for generator governor action, automatic voltage regulation, and protection system emergency responses which disconnect loads, lines, and generators when network parameters exceed pre-defined thresholds. 

\subsection{Power System Model}
\label{ssec:power system model}
We consider a third-order model for generator dynamics, which includes both frequency and voltage dynamics \cite{schm2014, goodridge2021}, enabling reliable estimation of frequency dynamics for longer transient periods. We model subsequent power grid emergency actions  occurring after the initial LAA, allowing us to capture critical actions during both primary and secondary frequency control over the entire simulation duration.

Consider a graph theoretic formulation of a power system $\mathcal{G} = ( \mathcal{N}, \mathcal{T})$, where $\mathcal{N}$ is the set of network buses and $\mathcal{T}$ is the set of transmission lines connecting the buses. The set $\mathcal{N}$ has cardinality $|\mathcal{N}| = N+L$, and contains $N$ generation buses and $L$ load buses. At each generation bus $i= 1, \dots N$, the dynamics for the voltage phase angle $\delta_i$, voltage magnitude $E_i$, and governor action $\rho_i$ are described by:

{\small\begin{subequations}
    \label{eq:network_model}
    \begin{empheq}[left = \empheqlbrace\,]{flalign}
            & M(\psi) \ddot{\delta}_{i}+D \dot{\delta}_{i}=\psi_i P^G_i- P^L_i(\iota_i)\nonumber \\ &\hspace{2.35cm} - E_{i}\sum_{j=1}^{N+L}B_{ij} \Omega_{ij}E_{j}\sin(\delta_{ij} )\label{eq:freq21} \\
            & S_{i}\dot{E}_{i}=\psi_i (E_{f,i}-\text{v}_i) -E_{i}+ X_{i}\sum_{j=1}^{N+L}B_{ij} \Omega_{ij}E_{j}\cos(\delta_{ij} )\label{eq:volt21}\\
            & \dot{\rho}_{i}= -A_{i}\dot{\delta}_{i}(1-1_{\mathcal{\mathcal{\mathcal{W}}}}[\dot{\delta}_{i}]).\label{eq:gov21}
        \end{empheq}
\end{subequations}}
\noindent In a similar manner, the dynamics for $\delta_i$ and $E_i$ at each load bus $i=N+1, \dots, N+L$ are given by: 
{\small \begin{subequations}
        \label{eq:network_model_load}
            \begin{empheq}[left = \empheqlbrace\,]{flalign}
                & M(\psi) \ddot{\delta}_{i}+D \dot{\delta}_{i}= -P^L_i(\iota_i)- E_{i}\sum_{j=1}^{N+L}B_{ij} \Omega_{ij}E_{j}\sin(\delta_{ij})\label{eq:freq_load} \\
                & S_{i}\dot{E}_{i}=\psi_i E_{f,i}-E_{i}+X_{i}\sum_{j=1}^{N+L}B_{ij} \Omega_{ij}E_{j}\cos(\delta_{ij}).\label{eq:volt_load}
        \end{empheq}%
\end{subequations}}%
In equations~\eqref{eq:network_model} and~\eqref{eq:network_model_load}, the indicator variables $\psi_i$, $\Omega_{ij}$ and $\iota_i$ represent the state of relays which control generator, line, and load disconnections respectively (discussed in~\cref{sec:emer_res}). \Cref{tb:model_variables} provides descriptions of model variables. 
\begin{table}[!h]
    \centering
    \caption{\small Variables used in~\eqref{eq:network_model} and~\eqref{eq:network_model_load}. }
    \begin{tabular}{|c|l|c|}
        \hline
        \textbf{Symbol}&\textbf{Meaning}&\textbf{Units}\\\hline
        $A_i$&Governor's droop response &MW/rad\\               
        $B_{ij} $&Susceptance matrix&p.u.\\   $P_i^G$&Net generation at node $i$&p.u.\\
        $P_i^L(\iota)$&Net loads at node $i$&p.u.\\
        $D$&System damping& \%\\
        $\delta_i$ & Phase angle at node $i$ & p.u\\
        $\delta_{ij}$&$\delta_i - \delta_j$&p.u.\\
        $\dot{\delta}_i$&Frequency at node $i$ & p.u\\
        $\ddot{\delta}_i$&Rate of change of frequency (RoCoF)&p.u.\\
        $E_i$ & Voltage at node $i$ &p.u.\\ 
        $E_{f,i}$&Machine $i$ rotor field voltage&p.u.\\         $M(\psi)$&System angular momentum & Ws$^2$\\
        $\Omega_{ij}$&Line $ij$  disconnection indicator&-\\
        $\psi_i$ & Generator shed indicator&-\\
        $\iota_i$&UFLS counter&-\\
        $S_i$&Machine $i$ transient time constant&s\\
        $X_i$&Machine $i$ equivalent reactance&ohms\\
        $\mathcal{W}$&Governor's deadband frequency range&Hz\\
        
        \hline
    \end{tabular}
    \label{tb:model_variables}
\end{table}

At nodes $i = [ 1, \dots N]$, net generation is given by $P^G_i \coloneqq \min\{P_i^{\text{max}},P^{e}_i + \rho_i\}$, where $P_i^{\text{max}}$ is the nominal maximum power output of generator $i$, $P_i^{e}$ is the equilibrium power of the generator at $t=0^-$ before any disturbance is applied to the network, and $\rho_i$ models the action of the generator governor, with dynamics given in \eqref{eq:gov21}. The variable v$_i$ accounts for automatic voltage regulation (cf.~\cite{goodridge2021} for further details). 


\subsection{Modelling network loads}
\label{sec: load model }
For each network, we model a fixed, maximum magnitude of loads $P_i^{TL}$ at each node. We model nominal equilibrium loads at $t =0^-$ as a fraction of total loads are modelled to be online and active $P_i^{eq} < P_i^{TL}$, reflecting typical power balances observed during critical points of the diurnal cycle. We introduce the network variable $\nu \in [0,1]$ to represent the proportion of total loads susceptible to LAAs (i.e., connectivity-enabled loads that are vulnerable to cyber attacks). Thus, we decompose the total loads at each node into a vulnerable component $\nu P_i^{TL}$, comprised of unsecured, IoT devices, and a secure component $(1-\nu)P_i^{TL}$, comprised of non-IoT or protected devices. 


{\bf Load-altering attack model:} 
LAA refers to the manipulation of the vulnerable component of network loads by an attacker to disrupt the power balance of the network  with the intention of leading the system to an unsafe state. Existing literature models two different types of LAAs  -- S-LAAs consisting of a one-time load manipulation) and D-LAAs consisting of multiple changes to loads in the network over a period of time \cite{Soltan2018, HuangUSENIX2019, AminiLAA2018, LakshIoT2021}.

In this work, we present a unified model for S- and D-LAAs consisting of \emph{discrete-time} load changes. In particular, we model load manipulations occurring at regular discrete times $t_j$ spaced $\mathcal{I}$ seconds apart, i.e., $t_j = (j-1) \mathcal{I}$ for $j =1, 2, \dots, n$, with $n$ load changes during the simulation duration denoted by $T_{\max}$ seconds. We assume that $\mathcal{I} \geq \mathcal{I}_{\min},$ where $\mathcal{I}_{\min} > 0$ represents the physical limits on how quickly load magnitudes can change. 
Under this model, an S-LAA can be considered a limiting case of a dynamic attack with $n=1$. Without a loss of generality, the first load change is modelled to occur at $t = t_1 =0$ for both strategies.

We model the $j^{th}$ \textit{commanded nodal load change} at time $t_j$ to be a time-dependent fraction of the equilibrium loads, namely $\eta_i(t_j) P_i^{eq}$, for some factor $\eta_i(t_j) \in [\eta_{\min},\eta_{\max}]$, where $\eta_i(t_j)<0$ ($\eta_i(t_j)>0$) can be understood as a command to reduce (respectively, increase) the load of node $i$ at time $t_j$\footnote{Note that this model also presents a generalized version of D-LAAs as compared to prior works \cite{AminiLAA2018, LakshIoT2021} which modelled them as \emph{continous} load changes and the load change pattern was restricted to being proportional to the instantaneous frequency deviations.}. The variables $\eta_{\min}$ and $\eta_{\max}$ represent the minimum and maximum allowed values of $\eta_i(t_j)$ in order to model the physical limits of the load change. For a fixed number $n$ of load changes, a load-altering attack is thus fully described by the $L \times n$ matrix $\eta$ whose entries describe all the commanded load changes, more specifically $\eta_{i,j} := \eta_i(t_j)$. The commanded load change at time $t$ at node $i$ can be modelled via the step-function $\hat{u}_i: \mathbb{R}^+ \mapsto \mathbb{R}^+$: 
\begin{equation}\label{eq:attackmagnitude}
    \hat{u}_i(t) = P_i^{eq} \sum_{j=1}^n \eta_i(t_j) \mathbbm{1}_{[t_j,t_{j+1})}(t),
\end{equation}
where $\mathbbm{1}_{[t_j,t_{j+1})}(t)=1$ if $t_j \leq t < t_{j+1}$ and $0$ otherwise. However, despite the commanded load change, the fixed quantity of loads on the network limits the realized load change experienced. Therefore, at time $t$ during the simulation, the actual net load at node $i$, including the realized LAA but not the emergency responses, is given by:


    


\begin{equation}       
    \hat{P}_i^L(t) \coloneqq (1-\nu)P_i^{eq} + \nu \min \big[0, \max\big(P_i^{TL}, P_i^{eq}+ \hat{u}_i(t)\big)\big].
\end{equation}
The second term on the RHS represents the realized load change, which restricts the authority of the attacker to manipulate the vulnerable loads present at node $i$ between a minimum of $0$~MW and an upper limit of $\nu P_i^{TL}$. This upper limit also highlights that the attacker can bring inactive vulnerable loads online during the LAA.
%
To record the realised load change applied at $t=t_j$, at each node, we define the set of time-dependent \textit{realized load changes} $\{\lambda_i(t_1),\lambda_i(t_2),\dots,\lambda_i(t_n)\}$, where each $\lambda_i(t_j) \in \mathbb{R}$ at node $i$ can be calculated as: 
\begin{equation}\label{eq:lambdatj}
\lambda_i(t_j) \coloneqq \hat{P}_i^L(t_j) - \hat{P}_i^L(t_j -\delta t), \quad j=1, \dots, n,
\end{equation}
where $\delta t$ represents a small time increment. At each node $i$, the realized load changes $\lambda_i(t_j)$'s can be aggregated over time to describe the full extent of the nodal attack, yielding the \textit{nodal cumulative realised attack} $\Sigma_i$ defined as 
\begin{equation}\label{eq:Sigma}
    \Sigma_i \coloneqq \sum_{j=1}^n \left|\lambda_i(t_j)\right|.
\end{equation}
This is a measure valid for both S- and D-LAAs of the total magnitude of realised load changes deployed against each node (in units of power) over the entire simulation.

\subsection{Emergency Responses}
\label{sec:emer_res}
We model independent network protection systems intended to emulate the systems designed to protect sensitive power system components from transients in network frequency and voltage following a change in the active power balance. We provide a brief description of these protection systems.

\subsubsection{Generation shedding} Two schemes are modelled to disconnect generators from the network: (1) excessive RoCoF induced generation shedding (RIGS), where generators are disconnected if the locally measured RoCoF $|\ddot{\delta}_i|$ exceeds a pre-specified upper threshold, and (2) over frequency generation shedding (OFGS), which disconnects generators when nodal frequency $\dot{\delta_i}$ deviations surpass a pre-specified upper threshold.


  
\subsubsection{Load shedding} Loads in the network can be shed via two independent schemes: (1) Under Voltage Load Shedding (UVLS) and (2) Under Frequency Load Shedding (UFLS). We model a simplified UVLS scheme that disconnects 5\% of nodal loads if nodal voltage magnitude $E_i$ falls below the pre-specified threshold of 0.9 p.u for more than 5 seconds. This mechanism activates only once and is designed to prevent voltage collapse. We introduce an indicator variable $\mathcal{V}_i$, which equals $1$ when the nodal voltage magnitude at node $i$ meets the activation criterion for the UVLS relay and $0$ otherwise.

The UFLS scheme is designed to prevent frequency collapse by progressively disconnecting loads when nodal frequency $\dot{\delta}_i$ falls below a strictly decreasing sequence of thresholds $F^U \coloneqq \{F^U_1,\dots,F^U_4\}$ where $F^U_{j-1} > F^U_j$ \cite{goodridge2021}. At each threshold, 10\% of equilibrium loads $P^{eq}_i$ is disconnected to arrest the decline in nodal frequency. Letting $\mathcal{F}_i \in \{0,1,2,3,4\}$ count the cumulative number of UFLS relay activations at node $i$ at each time step $t$ in the simulation, the net load inclusive of any UFLS and UVLS activations up to time $t$, is a dynamic variable in the power system model:		
\begin{equation}       
    P_i^L(t,\iota) = \iota_i \cdot  \hat{P}_i^L(t),
\end{equation}
where $\iota_i \coloneqq \big(1- 0.1\mathcal{F}_i - 0.05\mathcal{V}_i \big).$ 
Correspondingly, load-shedding events reduce the effectiveness of LAAs by disconnecting compromised loads, degrading the attacker's ability to further manipulate the network's power balance.
    
\subsubsection{Line disconnection} We model the disconnection of inter-area transmission lines, as their loss can exacerbate frequency instability and lead to islanding. When the power flow $\phi_{ij} \coloneqq B_{ij}E_i E_j \sin(\delta_{ij})$ through any interconnector line $(i,j)$ exceeds a pre-set power threshold $P^{\phi}$, the line $(i,j)$ is tripped, and power is no longer allowed to flow through it for the rest of the simulation. The indicator $\Omega_{ij}$ represents the state of the interconnector between nodes $i$ and $j$, with the value $1$ indicating nominal operation. Once the conditions for line disconnection are triggered, $\Omega_{ij}$ switches to $0$ for the rest of the simulation and the power flow on this line stops~\cite{goodridge2021}.

As the simulation evolves, the ER model inspects $\dot{\delta}_i(t)$, $\ddot{\delta}_i(t)$, $E_i(t)$ and $\phi_{ij}(t)$ from the power system model~\eqref{eq:network_model}. Once the activation criterion for an ER is observed, the corresponding ER is activated. This results in a discontinuity in \eqref{eq:network_model}, where changes to the relevant line, generator, or load are applied. The simulation is then resumed with the new network parameters.



\section{Statistical model and rare-event sampling approach for LAAs}\label{sc:statmodel}

Direct prediction of \emph{network-threatening} LAA characteristics is particularly challenging, as they can exploit unforeseen vulnerabilities in network architecture or operational practices, inducing emergency responses in unforeseen ways. Instead, this work employs a \textit{sampling} approach to identify LAAs that lead to at least one subsequent emergency response, by applying randomly generated load changes and network conditions (referred to as `proposals') to the power system described in~\cref{ssec:power system model}. We then construct the sample of successful LAAs by only retaining proposals that (1) trigger at least one emergency response and (2) follow a desired distribution. 

\subsection{The statistical model}
\label{ssec:stat_model}
To enable the aforementioned sampling approach, we account for the uncertainty in key LAA characteristics (magnitude and frequencies) and network parameters (load vulnerability and equilibrium conditions) as follows:

\begin{itemize}
\item \textit{Distribution of network variables:}
To explore the interplay between LAAs and different network operational configurations, we sample two network parameters: (i) the network vulnerability ratio $\nu$ from a uniform distribution $\nu \sim \mathcal{U}([0,1])$ and (ii) the diurnal power balance scenarios $\tau$ uniformly from the set $\{1,2,3,4\}$. These scenarios correspond to expected load magnitudes observed during night, morning, afternoon, and evening hours, respectively. The nodal equilibrium loads $P_i^{eq}$ under these load scenarios are proportional adjustments to the published power balance of the networks investigated \cite{ieee39reference, kundur1994power}. The relative factors $\{0.4, 1, 0.85, 1.3\}$ for $\tau=1,\dots,4$ reflect the peaks in the UK's daily load cycle \cite{ukdiurnalcycle2018}. 

\item \textit{Distribution of LAA frequency and magnitudes:}
For dynamic LAAs, we assume the time between attack $\mathcal{I}$ is a uniform random variable with $\mathcal{I} \sim \mathcal{U}([1, T_{\max}])$, excluding intervals of less than $1$ second to account for the time taken to process and execute programmed load changes (i.e., $\mathcal{I}_{\min} = 1$ second), and $T_{\max} = 60$ seconds is the total duration of each simulation. Once $\mathcal{I}$ has been sampled, we determine the corresponding number $n$ of attacks with $n = \lfloor T_{\max}/\mathcal{I} \rfloor$ and assume they are equally interspersed at times $t_j = (j-1) \mathcal{I}$ for $j=1,\dots,n$.
\item \textit{Distribution of LAA magnitudes:} We generate a set of commanded \textit{proportional} load changes by sampling uniformly at random from $[\eta_{\min},\eta_{\max}]^{L \times n}$ and storing them in a $L \times n$ matrix $\eta$. Each entry $\eta_{i,j}$ represents the load change factor at time $t_j$ with respect to the equilibrium load $i$. In particular, we set $\eta_{\min} = -1$ and $\eta_{\max} = 5$. The minimum value of $-1$ signifies reducing the entire vulnerable part of the load at node $i$ to zero, while positive values indicate load increases. The maximum load change factor is set to $5$ to account for the fixed total load magnitude in the network.
\end{itemize}

Therefore, in our setting, a load-altering attack is fully characterized by the quadruple $(\eta,\nu,\tau,\mathcal{I})$, where $\nu$ is the proportion of vulnerable loads, $\tau$ describes the network power equilibrium state, $\mathcal{I}$ is the interval between attacks and the $L \times n$ matrix $\eta$ whose entries describe all the commanded load changes (recall that $n$ is automatically determined by $\mathcal{I}$). Using the power system model (\textbf{PSM})~\eqref{eq:network_model}--\eqref{eq:network_model_load}, for each attack $(\eta,\nu,\tau,\mathcal{I})$, we calculate the nodal cumulative realized attack $\mathbf{\Sigma}$ and determine if it was successful $S$ by
\[
    (S, \mathbf{\Sigma}) = \textbf{PSM}(\eta,\nu,\tau,\mathcal{I}),
\]
where $S=1$ if the attack was successful and $0$ otherwise.

Ultimately, we desire the nodal LAAs in the final sample reflect a realistic distribution of LAA sizes expected in real-world attacks, where LAAs with larger cumulative magnitudes $\mathbf{\Sigma}$ are rare. To achieve this bias for each nodal LAA, we set a \textit{target probability density} $\rho$ on $\mathbb{R}^L_+$ to be 
$\rho \sim \prod_{i=1}^L \text{LogNormal}(\mu,\sigma^2)$, with parameters $\mu=1$ and $\sigma=5$. This joint distribution is supported on $[0,+\infty)^L$, and renders large magnitude LAAs at each node rare. 

Thus, letting $A$ denote the set (or `event') of LAA magnitudes that results in the activation of at least one emergency response, our goal is to sample from the conditional distribution
$
\pi = \frac{\rho(\mathbf{\Sigma})\mathbbm{1}_A}{\rho(A)}, 
$
where $\rho(\mathbf{\Sigma})$ is the probability density of an LAA with nodal magnitudes $\mathbf{\Sigma},$ (where $\mathbf{\Sigma} =(\Sigma_i)_{i=1}^L$), and $\rho(A)$ is the likelihood that an LAA triggers an emergency response. The indicator $\mathbbm{1}_A = S \in \{0,1\}$ takes a value of 1 if the LAA with nodal magnitudes $\Sigma$ is associated with the activation of at least one emergency response, and 0 otherwise. Together, $\pi$ describes a probability distribution conditioned on the activation of at least one emergency response occurring (event $A$).

\subsection{The Need for Rare-Event Sampling}
\label{ssec:sampling_procedure}

\begin{table*}[ht!]
    \begin{minipage}{\textwidth}
        \centering
        \caption{\small Comparison of sampling methodologies when drawing realizations of network threatening  nodal \textit{static}-LAAs, against two versions of the IEEE 39 test network -- one with protection systems and network characteristics tuned to satisfy the $N-1$ contingency criterion and a second one that is $N-0$ secure that does not consider contingencies in its design. Additionally, we investigate two uncertainty models for LAAs, the first, a uniform distribution (all LAAs are equally likely), and a LogNormal distribution (large magnitude LAAs are rare). All simulations were conducted for 100 hours, at which point the resulting sample size was recorded.}
        \begin{tabular}{|c|l|c|c|c|c|}
        \hline
        Network Security & Sampling Methodology & Underlying Distribution & Accepted samples & Acceptance Rate, \% & range $\nu$ \\ \hline \hline
        \multirow{3}*{$N-0$ Secure} & Monte Carlo & $ \mathrm{LogNormal}(1,5)$ & 0 & 0 & n/a \\ \cline{2-6}
         & Random walk Metropolis & $ \mathrm{LogNormal}(1,5)$ & 5000 & 5 & $[0.15, 0.99]$ \\ \cline{2-6}
         & \textbf{Skipping Sampler} & $ \mathrm{LogNormal}(1,5)$ & \textbf{57000} & \textbf{50} & $[0.12, 0.99]$ \\ \hline \hline
        \multirow{6}*{$N-1$ Secure} & Monte Carlo & $ \mathcal{U}[0, 5]$ & 0 & 0 & n/a \\ \cline{2-6}
         & Monte Carlo & $ \mathrm{LogNormal}(1,5)$ & 0 & 0 & n/a \\ \cline{2-6}
         & Random walk Metropolis & $ \mathcal{U}[0,5]$ & 200 & $0.0019$ & $[0.95, 0.99]$ \\ \cline{2-6}
         & Random walk Metropolis & $ \mathrm{LogNormal}(1,5)$ & 0 & 0 & n/a \\ \cline{2-6}
         & \textbf{Skipping Sampler} & $ \mathcal{U}[0,5]$ & \textbf{37000} & \textbf{37} & $[0.64, 0.99]$ \\ \cline{2-6}
         & \textbf{Skipping Sampler} & $ \mathrm{LogNormal}(1,5)$ & \textbf{9000} & \textbf{16} & $[0.66, 0.99]$ \\ \hline
        \end{tabular}
\label{tb: skipping vs MC vs RW}
    \end{minipage}
\end{table*}

\begin{figure}[t!]
    \centering
    \includegraphics[width=0.485\textwidth]{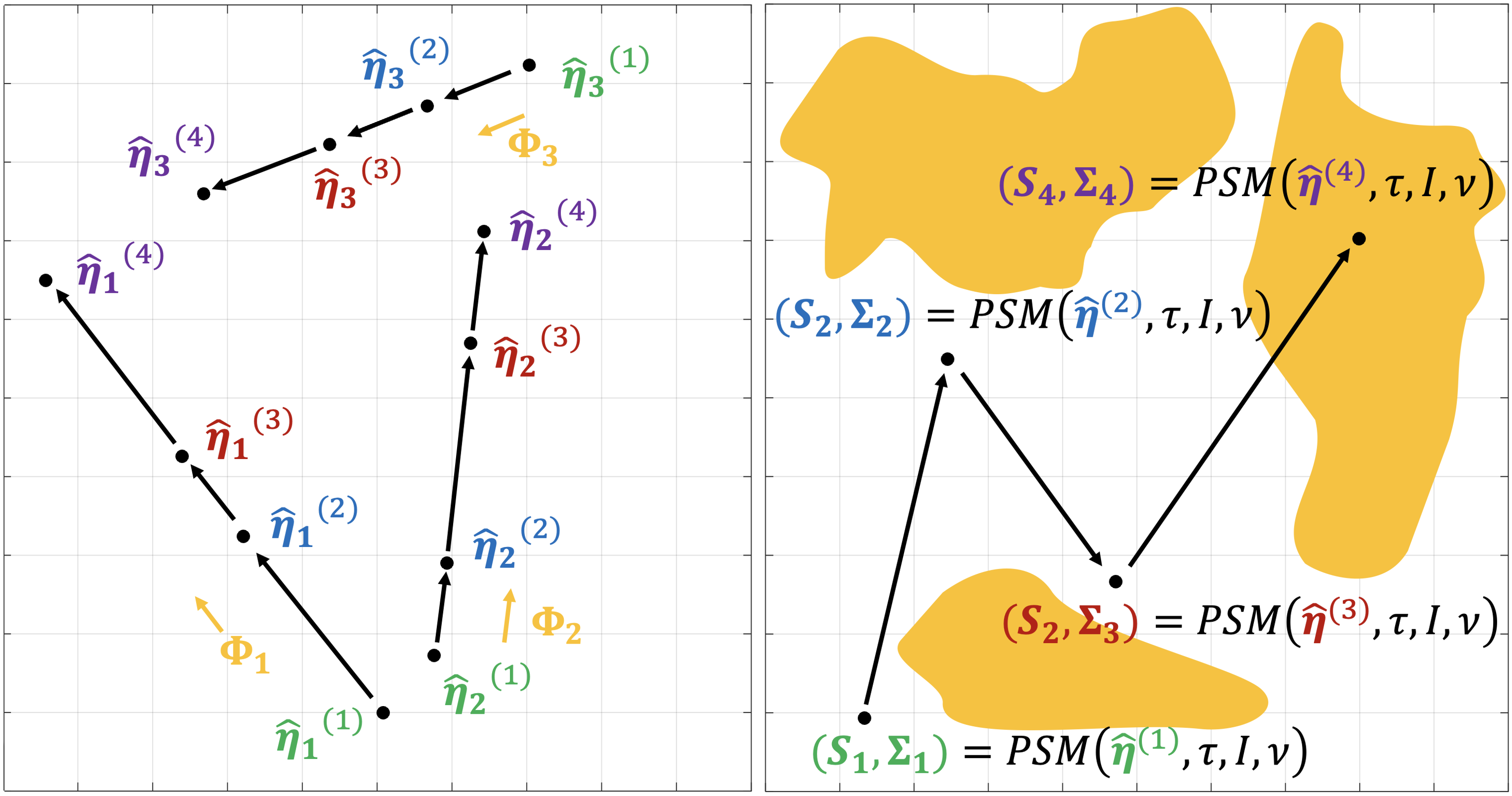}
    \caption{\small Illustration of a skipping proposal for a D-LAA in the commanded load change $\eta$ state space (where the `skipping' update occurs) [left] and the realised load change ($\Sigma$) state space (where proposals are accepted/rejected) [right]. The set $A$ of realised load changes associated with emergency responses is denoted in orange. Each sampled D-LAA is composed of $n = 3$ load changes in $L=2$ nodes and is thus described by three vectors $\hat{\eta}_1,\hat{\eta}_2, \hat{\eta}_3 \in \mathbb{R}^2$, which are thus represented as distinct points [left]. Instead of simply discarding the initial sampled unsuccessful LAA $\eta^{(1)}$ ($S_1=0$ and the corresponding point $\mathbf{\Sigma}_1$ is not in the target region $A$ in orange), each commanded load change (at $t_1$, $t_2$, and $t_3$) is updated in a linear fashion by randomly generating new points along the random directions $\Phi_1$, $\Phi_2$, and $\Phi_3$ respectively, determining $\hat{\eta}^{(2)}, \hat{\eta}^{(3)}, \dots$ and thus of $\mathbf{\Sigma}_2, \mathbf{\Sigma}_3,\dots$ (via the power system model, denoted $\textbf{PSM}$). This skipping process continues until the resulting realised load changes trigger at least one emergency response, corresponding to $\mathbf{\Sigma}_4$ (which is the first in the target region $A$ in orange, i.e., the first with $S_4=1$).} 
    \label{fig:skipping_example}
\end{figure}

Directly sampling $\pi$ using popular algorithms like \textit{Monte Carlo} samplers can be challenging due to several reasons: (1) Since networks are designed and operated to maximize resilience against contingencies, the event $A$ of interest is a rare event with a low probability density in realistic LAA distributions. Consequently, evaluating a large number of proposed attacks is necessary to identify a single relevant attack. (2) Understanding the characteristics of network-threatening LAAs requires sampling individual load changes at multiple nodes. For large networks, this leads to a high-dimensional sampling problem that can hinder algorithm performance. This problem is exacerbated for D-LAAs, which require sampling multiple load changes at each node. (3) Emergency responses depend on complex interactions between network conditions and the magnitude, spatial distribution, and timing of load changes. As a result, the set $A$ may exhibit complex geometry and possibly disconnected regions in the high-dimensional sample space of all possible LAAs (cf.~\cref{fig:skipping_example}).  

The results summarised in Table~\ref{tb: skipping vs MC vs RW} highlight the sampling challenge the statistical model presents. Consider the case of sampling network-threatening, static LAAs (S-LAAs) against 
the IEEE 39 test network.
A Monte Carlo sampler was unable to draw any samples of network-threatening LAAs in the stipulated time, regardless of network resilience or LAA uncertainty model. This is a consequence of the extreme rareness of such events in the high dimension sample space of the problem, leading the algorithm to spend the entire computational effort evaluating the more common, but irrelevant LAAs which do not threaten the network. 
This renders these approaches inefficient or ineffective when attempting to sample LAAs from $\pi$. 

Markov Chain Monte Carlo (MCMC) sampling techniques can overcome many of these challenges and enable efficient sampling of rare events. In general, MCMC algorithms can be summarised as a two-step procedure: (1) a \textit{Proposal step}, where a new, proposed state $Z$ for consideration is generated from a distribution of the user's choice (with few restrictions); (2) an \textit{Acceptance/rejection step}- using the target distribution $\pi$, the density of the proposed state $\pi(Z)$ is compared to the density of the most recently accepted state, denoted $Y$ (thus, $\pi(Y)$), always accepting $Z$ into the sample if its density is higher (i.e. it is more likely to occur than an already accepted state), and \textit{sometimes} accepting $Z$ if its density is lower. 

The generality of the MCMC methodology has led to multiple variants, including the well-known Random-Walk Metropolis (RWM) algorithm. However, standard formulations of RWM may also be ineffective when the event of interest (in our case, event \textit{A}) is both rare with respect to some underlying distribution and disconnected in the sample space. In these cases, RWM algorithms are liable to become `stuck' in one component of the event, unable to escape to draw samples from others \cite{Moriarty2019}. We observe evidence of this phenomenon when an RWM is applied to draw samples of network-threatening S-LAAs against different case studies involving the IEEE 39 network, cf.~\cref{tb: skipping vs MC vs RW}. The RWM algorithm generates valid S-LAA samples only in simplified sampling cases, where the set of threatening LAAs is much less rare in the state space, namely in the cases of: (1) an insecure $N-0$ IEEE 39 network with Log-Normal distribution for attack factors or (2) a $N-1$ secure IEEE 39 network with uniform distribution for attack factors. However, the limited range of $\nu$ in the sample suggests the algorithm became trapped in a local subset of $A$, and did not fully explore the state space. When applied to the desired, more challenging case of sampling threatening S-LAAs with a Log-Normal distribution against a ${N-1}$ secure network, the RWM also failed to generate any samples. Consequentially, these results justify the need for specialized algorithms to efficiently sample realistically distributed LAAs which may induce disconnections in large, ${N-1}$ secure power systems.

\subsection{Sampling Procedure}
The flexibility in the specification of the proposal step of MCMC algorithms allows users to devise variant methodologies to tackle unique sampling challenges. In this work, we apply one such variant, i.e., the skipping sampler that `skips' over $A^c$ until a point from $A$ is sampled or a specific halting index $K$ is reached (in our simulations, we use $K = 5$ for each skipping path).
We further amend this MCMC variant to provide a unified framework to generate both S-LAAs ($n =1$) and D-LAAs ($n > 1$) load changes 
leading to the methodology detailed in Algorithm~\ref{alg:skip}. In the special case of S-LAA, it is enough to consider the interval between attacks constant and equal to $T_{\max}$, which results in $n=1$ and a much smaller dimensional sampling problem. The main ideas underlying the procedure presented in Algorithm~\ref{alg:skip} can be summarised as follows:

{\small%
\begin{algorithm}[!h]
    \SetAlgoLined
    \SetKwInOut{Input}{Input}\SetKwInOut{Output}{Output}
    \textbf{Input:} halting index $K$, number of steps $m$;\\
    \BlankLine
    Sample a network state triplet as $(\nu, \tau, \mathcal{I}) \sim \mathcal{U}[0,1] \times \mathcal{U}[0,4] \times \mathcal{U}[1,T_{\max}]$;\\
    Calculate the number of load attacks $n = \lfloor T_{max} /\mathcal{I} \rfloor$;\\
    Generate a $L \times n$ matrix $\eta^{(0)}$ with i.i.d.~entries sampled from $\mathcal{U}[-1, 5]$ describing an initial set of commanded load changes;\\ 
    $(S_0, \mathbf{\Sigma}_0) = \textbf{PSM}(\eta^{(0)},\nu,\tau,\mathcal{I})$;\\
    $Y_0 \coloneqq [S_0, \mathbf{\Sigma}_0, \nu, \tau, \mathcal{I}]$;\\
    \For{$w = 0,1,\dots, m-1$}{ 
    Set $k=1$ and $\hat\eta^{(1)} \coloneqq \eta^{(t)}$; \\
    Generate $n$ skipping directions $\mathbf{\Phi}_j \in \mathbb{R}^L$ for $j = 1,\dots, n$;\\}
    \While{$S_k \neq 1 \textbf{\textup{ and }} k < K$}{
    Generate $n$ random distance increments $R_j^{(k)}$ conditionally on $\mathbf{\Phi}_j$ for $j = 1,\dots, n$;\\
        Calculate the new matrix of commanded load changes by setting 
        $\hat\eta^{(k+1)}_{i,j} \coloneqq \hat\eta^{(k)}_{i,j} + R_j^{(k)} \mathbf{\Phi}_j$ for $j = 1,\dots,n$;\\
        $(S_{k+1}, \mathbf{\Sigma}_{k+1}) = \textbf{PSM}(\hat\eta^{(k+1)},\nu,\tau,\mathcal{I});$\\
        Increment $k$ by $1$;\\
    }
    \uIf{$S_k=1$}{
    Set $Z \coloneqq [S_k, \mathbf{\Sigma}_k,\nu,\tau,\mathcal{I}]$;\\
    Evaluate the acceptance probability:
    \begin{equation*}
        \alpha(\ensuremath{Y_t},\ensuremath{Z})=
        \begin{cases}
            \min\left(1, \frac{\pi(\ensuremath{Z})}{\pi(\ensuremath{Y_t})} \right) & \text{ if } \pi(\ensuremath{Y_t}) \neq 0, \\
            1, & \text{ otherwise, }
        \end{cases}
    \end{equation*}
    Generate a uniform r.v.~$V$ on $(0,1)$;\\
    \uIf{$V \leq \alpha(\ensuremath{Y_i},\ensuremath{Z})$}{$\ensuremath{Y}_{w+1}=\ensuremath{Z}$;}
    \Else{$\ensuremath{Y}_{w+1}=\ensuremath{Y_w}$;}
    }
    \Else{$\ensuremath{Y}_{w+1}=\ensuremath{Y_w}$;}
    \textbf{Output:} $[Y_1, Y_2, \dots, Y_m]$\\
    \caption{Adjusted skipping sampler algorithm}
    \label{alg:skip}
\end{algorithm}}

\textbf{The proposal step}: We first sample (i) the parameters governing the network, i.e., vulnerability ratio ($\nu$) and the load scenario ($\tau$), and (ii) the interval between attacks ($\mathcal{I}$), which to compute the number of load changes as $n = \lfloor T_{max} /\mathcal{I} \rfloor$. We then generate $n$ distinct $L$-dimensional commanded load changes $\eta^{(0)}_1, \dots \eta^{(0)}_n$ uniformly at random from $[-1,5]^{L}$ (recall that $\eta^{(0)}_j$ encodes the attack at time $t_j$). 
These, along with the network parameters, are evaluated in the power system model. 
If no emergency response is activated, we return to and update each commanded load change using the key feature of the skipping sampler- we linearly displace and thus update each $\eta^{(0)}_j \to \eta^{(1)}_j = \eta^{(0)}_j + \Phi_j r_j$ by generating $n$ trajectories $\Phi_1,\dots,\Phi_n$ from a unit Gaussian density as well as $n$ random distance increments $r_1,\dots, r_j$ from a suitable distribution with strictly positive support. The updated load changes $\eta^{(1)}_1, \dots \eta^{(1)}_n$, along with the previous network conditions, are applied to the power system as before. This linear update for $\eta^{(k)}_j$ is repeated by sampling further random distance increments until either an emergency response is reported by the power system model, or the halting index $K$     is reached. Taking the final outcome of this subroutine, the resulting cumulative LAA magnitudes $\Sigma_k$, emergency response indicator $S_k$ as well as the network parameters ($\nu, \tau, \mathcal{I}$) are reported as the proposal $Z$ to be accepted or rejected in the next step. 

\textbf{The acceptance/rejection step}: The proposal $Z\coloneqq (\Sigma_k,\nu,\tau,\mathcal{I})$ is evaluated using the target density $\pi$ to determine whether it should be admitted to the final sample. As described earlier for all MCMC algorithms, if the density of $Z$ ($\pi(Z)$) is greater than the density of the most recently accepted state $\pi(Y_w)$, then $Z$ is added to the final sample and can be understood as the algorithm moving to a more likely state in the density $\pi$. Conversely, if its density is lower, it is sometimes accepted (as controlled by the uniform r.v.~$V$). This move to a less desirable state allows the algorithm to escape high-density regions of the state space and promotes greater exploration.  

This particular formulation of an MCMC algorithm allows more efficient sampling of rare events which may be disconnected in state space, as the skipping proposal mechanism prioritizes exploration of the state space. This emphasis allows the algorithm to avoid being trapped in local modes of $\pi$ depending on the initial starting point of the algorithm, a potential weakness of other MCMC algorithms (e.g. a random walk Metropolis algorithm). Further, the use of multiple skipping paths for each load change over time offers both a general and practical mathematical framework to represent both S-LAAs and D-LAAs. This technique 
eliminates the need to change the dimensions of the state space between proposals when more (or fewer) load changes are required.

As per~\cref{tb: skipping vs MC vs RW}, in contrast to MC and RMW algorithms, the skipping approach generated sufficiently large samples in all cases and was the only algorithm to generate S-LAAs of interest in the most realistic scenario where a resilient, $N-1$ compliant network was subjected to LAAs with rare large attacks (LogNormal distribution), thus showing the effectiveness of the methodology in sampling relevant LAAs.


%
%

\section{Simulations}\label{sec:Sims}




\subsection{Simulation Settings}
\label{sc:ktas}
We consider two well-known test networks in the literature: the Kundur 4-bus, two-area system (KTAS) \cite{kundur1994power} and the IEEE 39 network \cite{ieee39reference}. The Kron-reduced versions of these networks are used, with the reduced KTAS comprising $N=4$ nodes with generation units and $L=2$ pure loads and the reduced IEEE 39 network consisting of $N=10$ generator buses (two of which are modelled to include loads as well), and $L=17$ pure load buses. In the KTAS network, the interconnectors correspond to the transmission lines connecting Areas 1 and 2 (see \cite{GoodridgePMAPS2022}). The larger IEEE 39 network was divided into 3 areas, connected by a total of 4 interconnector lines (see \cite{ieee39reference}). The emergency response parameters are adjusted to ensure $N-1$ security for  each system.  



Simulations are conducted independently for each network and for each LAA type (dynamic and static). The system is initialized in one of 4 power equilibrium states specified by the sampled load scenario at $t=0^-$. The initial states of each system\footnote{Note: Parameters for each network can be found at \url{https://github.com/maldongoodridge/power_system_model_MG.git}} variable ($\delta_i\left(0\right)$, $E_i\left(0\right)$,  $\rho_i(0)$) are determined numerically such that $\ddot{\delta}_i \approx 0$.
The skipping sampler is employed to generate a sample of LAA magnitudes and network parameters, conditioned on at least one activation of an emergency response. In each proposal step of the algorithm, we generate a matrix of commanded load changes $\hat{\eta}^{(k)} \in \mathbb{R}^{L \times n}$. Along with the network variables, these inputs are fed into the power system model~\eqref{eq:network_model}, which we use to simulate frequency and voltage dynamics for $T_{\max}=60$ seconds using MATLAB. For each experiment, we conducted $50,000$ proposals.


The proposal densities of the sampling procedures are tuned separately for IEEE 39 and KTAS simulations to achieve an acceptance rate (the proportion of accepted proposals with respect to the total number of proposals) between 10\% and 40\%, as recommended in the literature \cite{Gamerman2006} for adequate exploration of the sample space. This results in sample sizes of $\sim$14000 for D-LAAs, and $\sim$9000 for S-LAAs, providing sufficient data for meaningful inferences. 

\vspace{-0.4 cm}

\subsection{Comparison of S- and D-LAAs via Statistical Analysis} 
\begin{figure}[!t]
    \vspace*{-0.25cm}
    \centering
    \subfloat[][Distribution of $\mu_{\lambda^-}$ for high-frequency D-LAAs ($\mathcal{I} < 30$s).]{\includegraphics[width=0.5\textwidth]{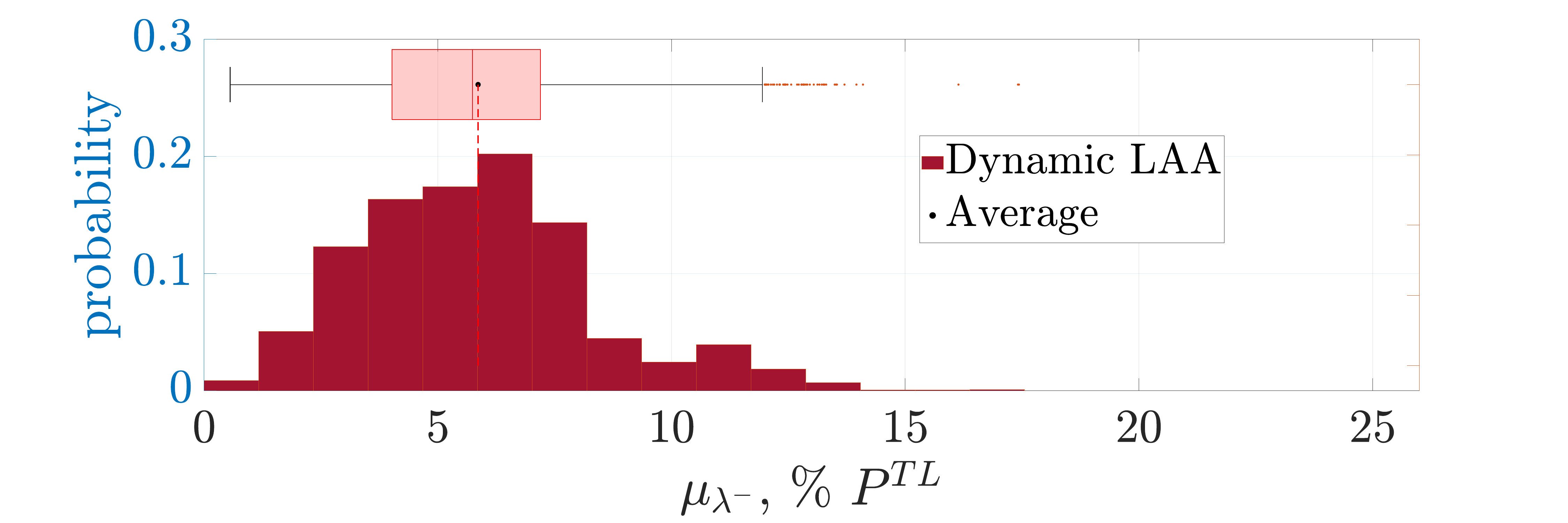}\label{fig: IEEE39 dyn mu lambda hflaa} }\\
    \subfloat[][Distribution of $\mu_{\lambda^-}$ for low-frequency D-LAAs ($\mathcal{I} \ge 30$s).]{\includegraphics[width=0.5\textwidth]{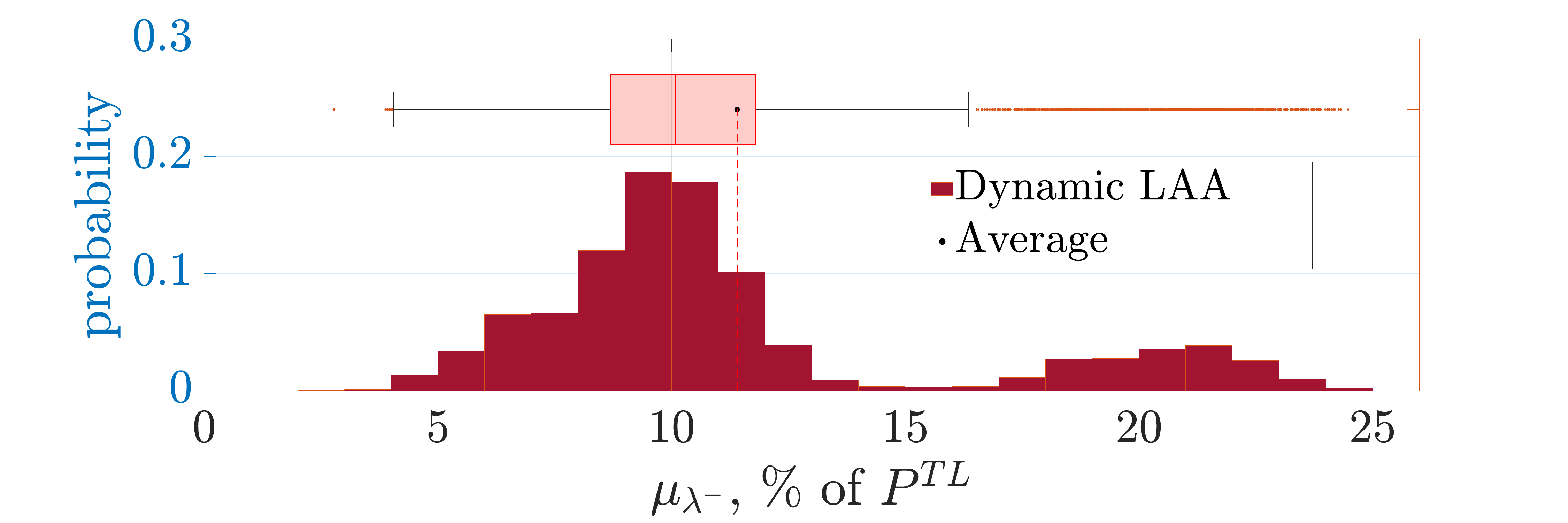}\label{fig: IEEE39 dyn mu lambda lflaa}}\\
    
    \subfloat[][Distribution of $\mu_{\lambda^-}$ for S-LAAs .]{\includegraphics[width=0.5\textwidth]{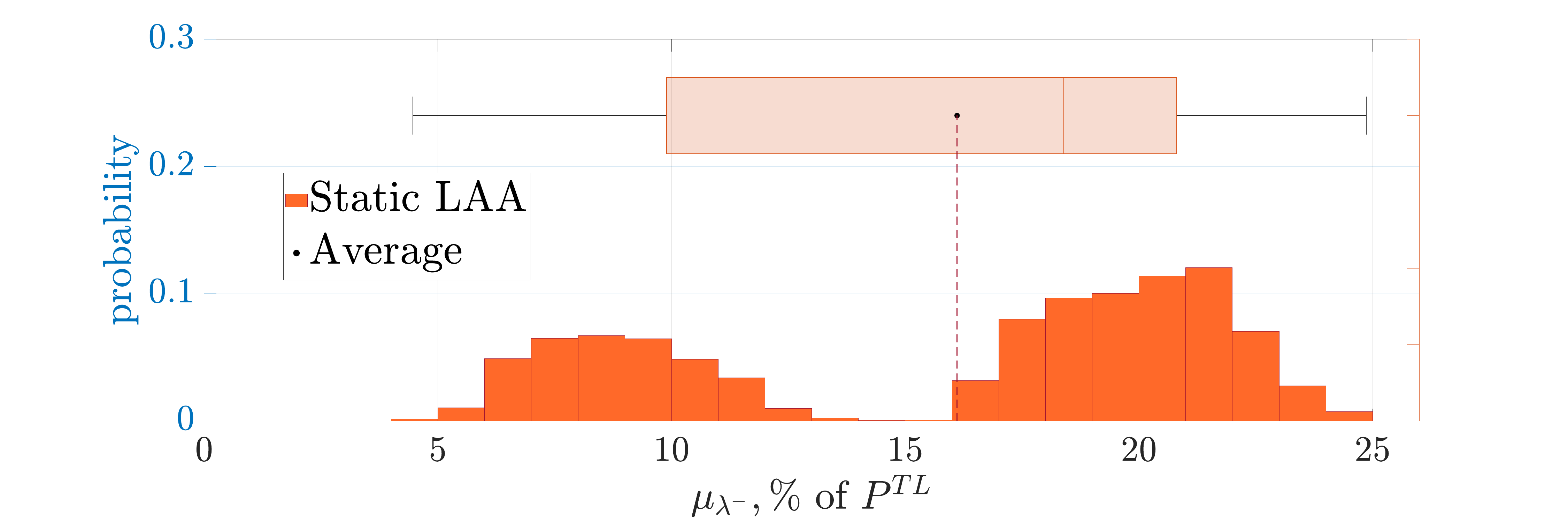}\label{fig: IEEE39 sta mu}}

    \caption{\small Distributions of the average load change magnitude during response time $\mu_{\lambda^-}$ for D-LAAs and S-LAAs applied against the IEEE 39 network, conditioned on the occurrence of at least one disconnection event $X$. Illustrations are provided for high-frequency D-LAAs, low-frequency D-LAAs, and static LAAs.} \label{fig: IEEE39 mu static v dyn}
    
\end{figure}

We perform statistical analysis (of the impactful LAAs uncovered by the skipping sampler in Section III-B) to compare S-LAAs and D-LAAs in terms of the average magnitude of  load change \textit{before} the first disconnection event, denoted $\mu_{\lambda^-}:= \sum_{i=1}^{L} \int_0^{r}\lambda_i(t_j)\mathbbm{1}_{\{t=t_j\}} dt$ where $t = r$ is the \textit{response time} -- the time of first disconnection for S-LAAs and D-LAAs. 
Simulation results for the IEEE 39 network (\cref{fig: IEEE39 mu static v dyn}) reveal that both the average and minimum of $\mu_{\lambda^-}$ are lower for D-LAAs (\cref{fig: IEEE39 dyn mu lambda hflaa}) as compared to S-LAAs (\cref{fig: IEEE39 sta mu}), with a hypothesis test comparing the averages revealing the difference, is statistically significant ($p$-value = $2\times 10^{-25}$). The reason behind this is that static attacks must deliver the entire load change needed to destabilize the network in a single event at time $t = 0$, leading to a bias towards larger magnitude load changes. Conversely, the ability to induce multiple load changes provides additional mechanisms to trigger disconnections by varying the temporal characteristics of the LAA. This reduces the reliance on magnitude-dominant approaches in LAA design and enables the use of lower-magnitude D-LAA strategies. 




		
In the IEEE 39 network, we observe bi-modality of the conditional distribution of $\mu_{\lambda^-}$ for both low-frequency D-LAA and S-LAA strategies. These modes correspond to different attack strategies during varying load balance scenarios ($\tau$). The lower mode is associated with LAAs leading to emergency responses during the peak demand period (\textit{Evening}), while the higher mode is linked to LAAs during the nadir demand period (\textit{Night}). Further details explaining these observations are provided in~\cref{ssec: impact of scenario}. In contrast, the distribution of $\mu_{\lambda^-}$ for high-frequency LAAs lacks bi-modality, indicating the inherent threat posed by this attack strategy to network operations. Such attacks induce disconnections in all scenarios with load changes of much smaller average magnitude.

\vspace{-0.4 cm}

\subsection{Impact of Intra-Day Power Equilibrium}
\label{ssec: impact of scenario}
\begin{figure}[!t]
    \hspace{-2.1cm}
    {\includegraphics[width=0.72\textwidth]{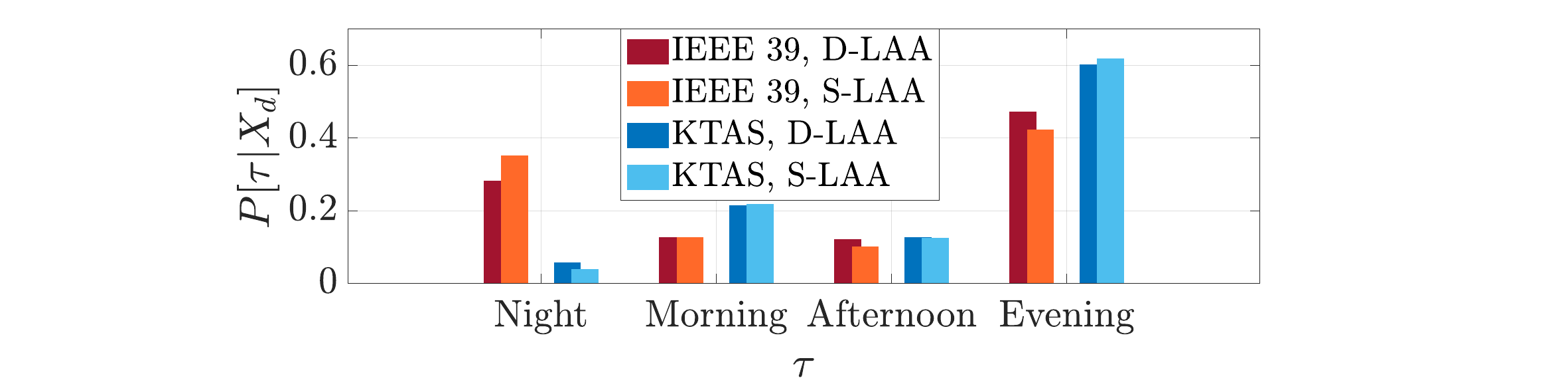}}\vspace*{-0.05cm}
    \caption{\small Average cascade size for the IEEE 39-bus network.}
    \label{fig: tod_prob_graph}
\end{figure}
We examined the vulnerability of both networks to LAAs under different load balance scenarios, reflecting diurnal demand cycles. In the KTAS network (\Cref{fig: tod_prob_graph}), the risk of disconnections is proportional to the magnitude of equilibrium loads. This risk is maximum during periods of peak demand when adversaries can induce a disconnection event by manipulating only $\sim 6\%$ of equilibrium loads (see \cref{tb:tod_table}). This low threshold is due to the larger absolute magnitude of loads online during peak demand and the network's high-stress state in this period. Thus, even marginal increases in loads can overwhelm network generation and trigger a localized load-shedding event. 
Conversely, the risk of LAA-induced disconnections is minimized during nadir demand periods (referred to as \textit{Night}). Although disconnections are rare during this scenario, our analysis provides insights into how such events could occur. In this scenario, successful attacks strategies are power-dominated instead of period-dominated: observing $\mu_{\lambda^-}$ in~\cref{tb:tod_table}, both S-LAA and D-LAA strategies require attackers to manipulate large proportions of active loads to trigger disconnections. While an attacker can induce disconnections using D-LAAs with smaller load change magnitudes, this dramatically extends the response $r$, underscoring the dominance of low-frequency, large-magnitude strategies in this scenario. 


\begin{table*}[h!]
    \centering
    
    \caption{\small Summary of successful attacks in terms of the average magnitude of each load change ($\mu_{\lambda}$), average response time $r$, and average attack period $\mathcal{I}$ for each network and LAA strategy.}
    \begin{tabular}{|c|c|c|c|c|c|c|}
        
        \hline
        \textbf{Network}& LAA & Metric &Night & Morning & Afternoon & Evening \\ \hline
        \multirow{5}*{IEEE 39} & \multirow{3}*{D-LAA}& Avg. $\mu_{\lambda}$ & 35.70 & 24.35 & 31.37 & 16.07\\ \cline{3-7}
        & & Avg. $r$, s & 34.75 & 48.82 & 39.99 &50.79\\ \cline{3-7}
        &&Avg. $\mathcal{I}$, s&32.25&31.45&30.46&33.96\\ \cline{2-7}
        & \multirow{2}*{S-LAA} & Avg. $\mu_{\lambda}$ & 44.1& 43.91 & 48.96 & 29.69\\ \cline{3-7}
        && Avg. $r$, s & 29.60 &48.62 & 40.73 & 50.17 \\ \hline
        \multirow{5}*{KTAS} & \multirow{3}*{D -LAA} & Avg. $\mu_{\lambda}$ &55.29&18.86&32.62 &6.31\\\cline{3-7}
        & & $r$, s& 18.37 &   8.28 &  10.93   & 8.39\\\cline{3-7}
        &&Avg. $\mathcal{I}$,s&30.33&31.89&31.11&31.63\\\cline{2-7}
        & \multirow{2}*{S-LAA}& Avg.$\mu_{\lambda}$&72.36&20.17&36.78&6.56\\\cline{3-7}
        
        & & $r$, s&5.88  &  7.23 &   6.64 &   8.09\\\hline
    \end{tabular}
    \label{tb:tod_table}
\end{table*}

LAA-induced emergency responses exhibit fundamental differences in the larger and more connected IEEE 39 network. Firstly, successful attack strategies in this network require multiple load changes, with the average response time $r$ exceeding the average attack period $\mathcal{I}$ as shown in~\cref{tb:tod_table}. This contrasts with the KTAS network, where an initial load change at $t=0$ was typically sufficient to induce the first disconnection event. Secondly, the results highlight how successful LAA strategies in the IEEE 39 network depend on the load scenario. During low-demand scenarios, larger magnitude load changes are required on average to trigger disconnections, resulting in a significantly reduced response time. Conversely, during peak demand periods, the preferred strategy is to manipulate a smaller proportion of loads to trigger a disconnection, providing more response time.  

\begin{figure}[!t]

    \subfloat[][Average cascade sizes during nadir demand \textit{Night} scenario.]{\includegraphics[width=0.48\textwidth]{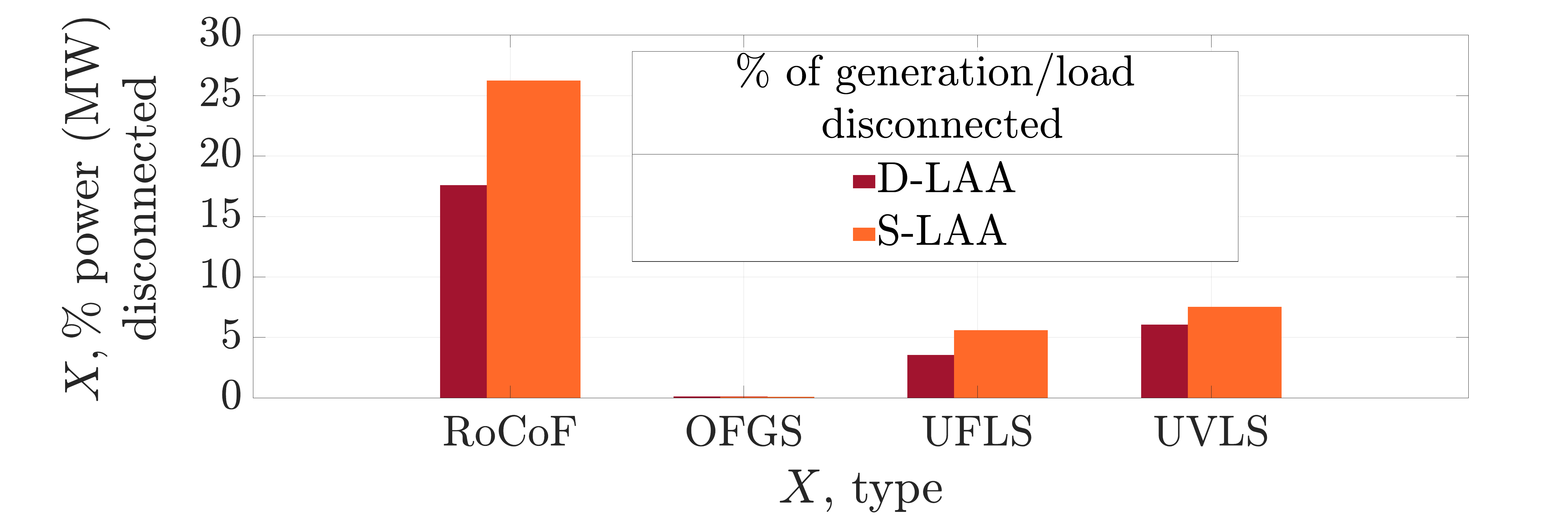}\label{fig: IEEE39 cascade morn}} \\
    
    \subfloat[][Average cascade sizes during peak-demand \textit{Evening} scenario.]{\includegraphics[width=0.48\textwidth]{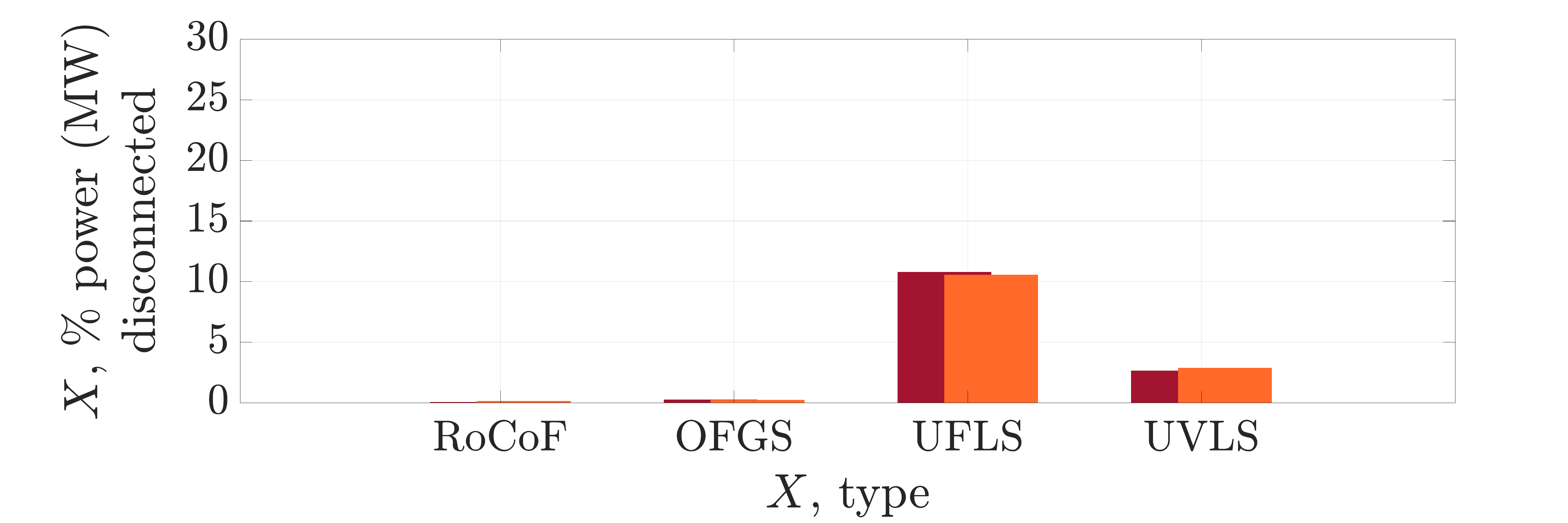}\label{fig: IEEE39 cascade even}} 
    
    \caption{Average cascade size for the IEEE 39-bus network.} 
    \label{fig:IEEE39 casade}
\end{figure}

These distinct characteristics of successful attack strategies based on the load scenario are also evident in the induced emergency responses, see~\cref{fig:IEEE39 casade}. During low demand periods, RoCoF generation shedding dominated power disconnected over the entire simulation, leading to the loss of a significant proportion of generation in the network. This highlights the network's vulnerability to power-focused LAA strategies during the nadir demand period (\textit{Night}). 
In higher demand scenarios, the damping effect of loads, coupled with the relatively smaller magnitude of load changes, limits RoCoF effects. Instead, successful LAAs in this regime exploit the cumulative effect of multiple load changes to gradually overwhelm already stressed generation capabilities, resulting in UFLS dominating emergency responses. Generation losses, a common failure mechanism profile of network overload, are rare in this regime.

Overall, these findings suggest the IEEE 39 network is particularly vulnerable to LAA-induced disconnections in periods of extreme demand. This is further supported by the bi-modal distribution of disconnections conditioned on load scenarios in~\cref{fig: tod_prob_graph}, with increased vulnerability in nadir (Night) and peak (Evening) demand scenarios.

\hspace{-1.5cm}



\subsection{Analysis of D-LAA Attack Patterns} 
\vspace{-0.2 cm}
In the literature, a common descriptor of D-LAAs which potentially threaten network security is the so-called `reverse governor' model, characterized by a negative relationship between network frequency and the effective network load change. Such load changes act in opposition to and therefore exacerbate frequency deviations, eventually exceeding preset protection systems thresholds for activation \cite{AminiLAA2018}.

We use a simple linear model, $\hat{\Sigma} = \hat{\beta}_0 - \hat{\beta}_1 \dot{\underline{\delta}}$, where $\hat{\Sigma} \coloneqq \sum_{i=1}^L\lambda_i^k$ is the net load change over the network, $\dot{\underline{\delta}}$ is the system frequency at the time of each load change event, and $\beta_1$ is the slope coefficient measuring the impact of load changes on frequency.
The reverse governor model posits that $\beta_1$ should be negative. We investigate this claim using the samples generated from our simulations.
\begin{figure}[!t]
    \vspace*{-0.3cm}
    \hspace*{0.2cm}
    {\includegraphics[width=0.4\textwidth]{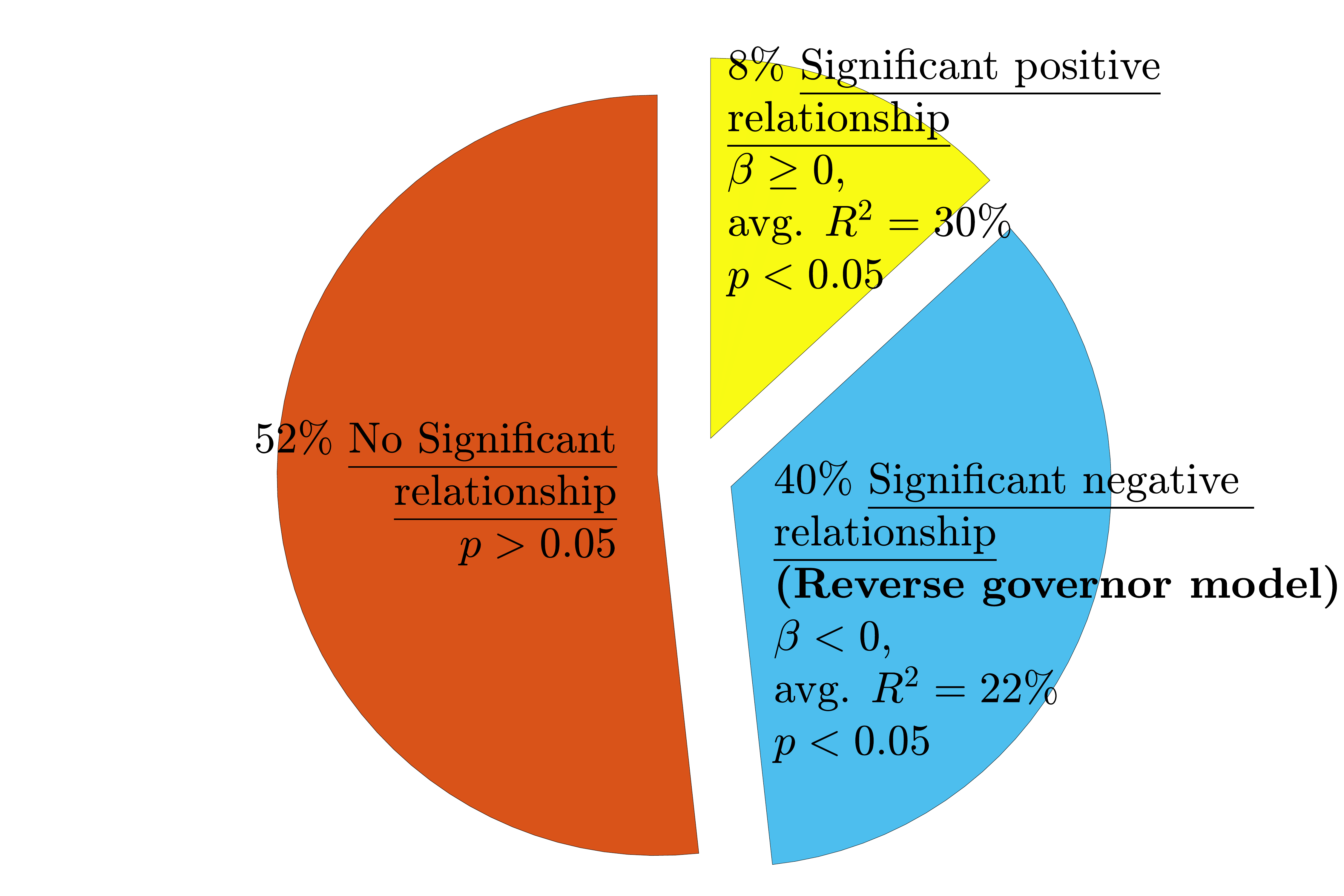}} 
    
    \caption{\small Summary of linear regression analyses between network frequency and network load changes in each sample. The chart illustrates the proportion of results where the relationship between load changes and frequency conforms to the reverse governor model (blue), the proportion which does not (yellow), and the proportion where there is no statistically significant relationship between loads and frequency (red). Samples with fewer than 3 load changes were excluded to remove trivial regression cases. We take a $p$-value less than 5\% as an indicator of statistical significance.}\label{fig: b1 chart}
    
\end{figure}
The pie-chart in~\Cref{fig: b1 chart} suggests a reverse governor model was statistically valid in only 40\% of cases sampled where LAAs led to subsequent emergency responses. In these cases, negative slope estimators ($\hat{\beta} < 0$) with p-values less than $0.05$ indicate there is less than a 5\% chance this relationship is spurious. However, in 52\% of the sample, there was no statistically significant relationship between network frequency and LAA load changes. Surprisingly, in 8\% of samples, there was even a positive relationship between load changes and frequency deviations, directly contradicting the assumption of a reverse governor model for D-LAAs. 

These results reflect the literature in that when a meaningful relationship between load changes and frequency can be extrapolated, the reverse governor model dominates, describing $80\%$ of such cases. However, in general, the reverse governor model is applicable only to a minority of all potential failure scenarios. The statistical analysis highlights variations in network frequency alone explained 22--30\% of the variation in network loads, as measured by the average coefficient of determination $R^2$. This suggests the reverse governor model is actually only one of many potential dynamic attack strategies.
\vspace{-0.2 cm}
\section{Conclusions and Future Research}
\label{sec:Conc}
In this work, we present a rare-event sampling framework to uncover impactful LAAs in power grids designed with $N-1$ security criteria and provide a rigorous comparison between S-LAAs and D-LAAs. Our results highlight the importance of adopting specialized sampling methodologies such as the skipping sampler in discovering impactful LAAs. Furthermore, we draw the following conclusions: (i) D-LAAs can induce power grid failures by manipulating a smaller fraction of IoT loads (cumulatively) as compared to S-LAAs. (ii) In low-demand scenarios, larger magnitude load changes are required on average to trigger disconnections (i.e., the dominant attack strategy being S-LAAs). Conversely, during peak-demand periods, manipulation of a smaller and high-frequency load manipulation triggers a disconnection (i.e., the dominant attack strategy being D-LAAs). These findings are corroborated by the time taken and the nature of emergency responses. (iii) We discovered a significant number of attack samples that do not conform to the `reverse governor' attack model proposed in \cite{AminiLAA2018}, highlighting the need for advanced sampling methodologies to identify impactful LAAs. In the future, we will investigate defence measures to enhance the resilience of power systems to LAAs using the insight derived from this work.

\vspace{-0.3 cm}
\bibliographystyle{IEEEtran}
\bibliography{IEEEabrv,bibliography}

\vspace{-0.3 cm}
\section*{Appendix: Case Studies}
We present two case studies below, one which exemplifies the reverse governor LAA model, and one which does not.
\begin{figure}[!h]
\vspace*{-0.3cm}
\hspace*{-0.3cm}
\centering
\subfloat[][Example of a high-frequency D-LAA which exhibits a `reverse governor' relationship between frequency and load changes.]{\includegraphics[width=0.45\textwidth]{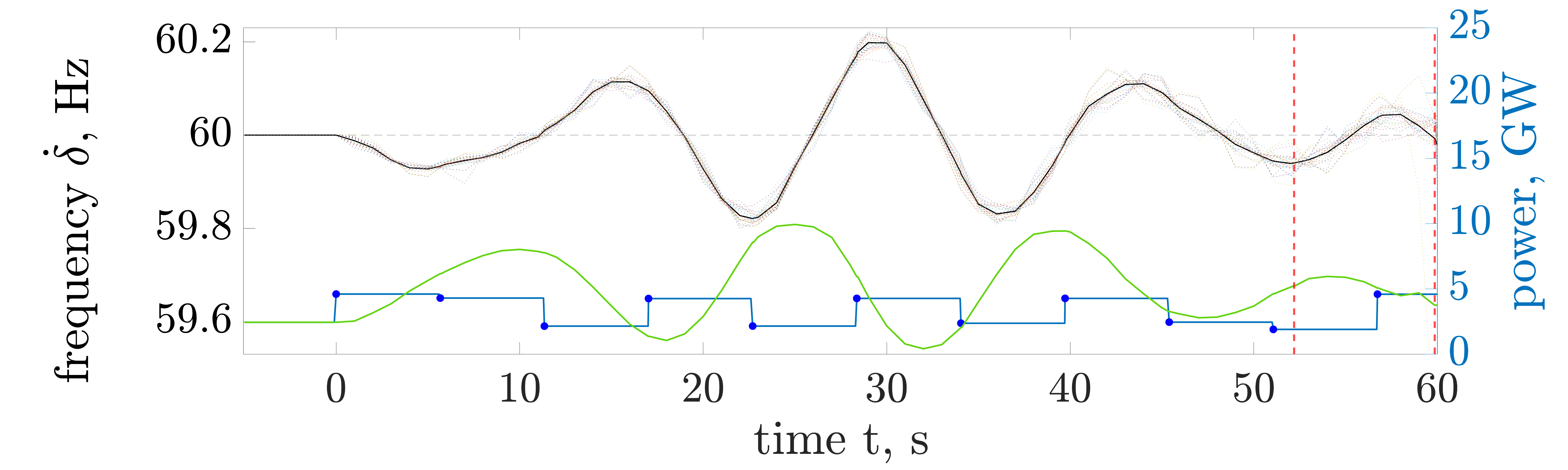}\label{fig: ieee_case_study_hflaa}} \\
\hspace*{-0.3cm}		
\subfloat[][Example of a low-frequency D-LAA where there is no significant relationship between loads and frequency.]{\includegraphics[width=0.45\textwidth]{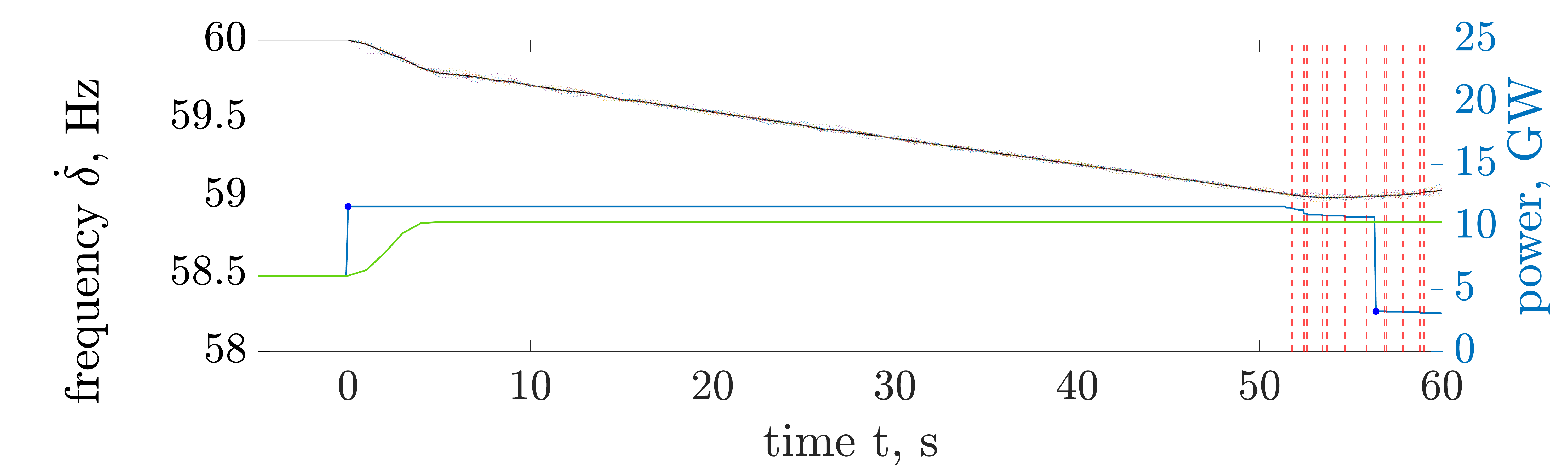}\label{fig: ieee_case_study_lflaa}}\\

\caption{\small Network frequency profiles (black, with nodal frequencies super-imposed in colour), network generation profiles (green), and network load profiles (blue) during simulated D-LAA in the IEEE 39 network. Load changes are depicted with a blue marker and emergency responses as vertical red dashed lines at the time of their occurrence.}\label{fig: ieee_case_study}
\end{figure}

\Cref{fig: ieee_case_study_hflaa,fig: ieee_case_study_lflaa} displays frequency, power, and load dynamics during simulated D-LAAs against the IEEE 39 network. At $t=0^-$, the system is in equilibrium, with loads equal to generation and network frequency of 60Hz. \Cref{fig: ieee_case_study_hflaa} illustrates a high-frequency D-LAA where load changes oscillate in opposition to generation to exacerbate both frequency deviations and RoCoF, eventually triggering a loss of load event and the disconnection of a generator at 51 and 59 seconds respectively. This attack was modelled to occur in the low-demand night scenario, thus generators had excess capacity to attempt to stabilize frequency excursions. However, the adversary exploited this response mechanism by commanding alternating load changes which exacerbate the frequency dynamics instead of dampening them.
   
In~\cref{fig: ieee_case_study_lflaa}, the D-LAA is instead deployed against the IEEE 39 network during a peak loading period when generation is near capacity. The attacker deploys a low-frequency D-LAA strategy, where at $t=0$, the initial attack increases loads beyond the capacity of the online generation. Thus, even after ramping to full capacity, there is net demand in the network, leading to a collapse in frequency. With no recourse, under frequency load shed relays independently activate across the network to arrest the decline in frequency. As the failure sequence is primarily induced by the initial LAA, there is little relationship between load changes and frequency deviations, violating the tenets of the reverse governor model. 

\end{document}